\documentclass[
12pt,
tightenlines,
aps,
nofootinbib,
floatfix]{revtex4}

\usepackage{graphicx}
\usepackage{float}
\usepackage{amsmath}

\usepackage{color}

\newcommand{\be}{\begin{equation}}
\newcommand{\ee}{\end{equation}}
\newcommand{\ba}{\begin{eqnarray}}
\newcommand{\ea}{\end{eqnarray}}

\def\vec#1{{\mbox{\boldmath$#1$}}}

\newcommand{\p}{\mbox{$\vec{p}$}}

\begin{document}
\begin{titlepage}

\begin{flushright}
\vbox{
\begin{tabular}{l}
ANL-HEP-PR-13-42\\
\end{tabular}
}
\end{flushright}
\vspace{0.1cm}


\title{
 ${ t\bar{t}}$+large missing energy from top-quark partners: a comprehensive study at NLO QCD
}

\author{Radja Boughezal}
\email[]{rboughezal@anl.gov}
\affiliation{High Energy Physics Division, Argonne National Laboratory, Argonne, IL 60439, USA}

\author{Markus Schulze}
\email[]{markus.schulze@anl.gov}
\affiliation{High Energy Physics Division, Argonne National Laboratory, Argonne, IL 60439, USA} 

\begin{abstract}
\vspace{2mm}

We perform a detailed study of top-quark partner production in the ${t}\bar{{t}}$ plus large missing energy final-state at the LHC,  
presenting results for both scalar and fermionic top-quark partners in the semi-leptonic and dileptonic decay modes of the top quarks.
We compare the results of several simulation tools: leading-order matrix elements, next-to-leading order matrix elements, leading-order plus parton shower simulations, and merged samples that contain the signal process with an additional hard jet radiated.   
We find that predictions from leading-order plus parton shower simulations can significantly deviate from
NLO QCD or LO merged samples and do not correctly model the kinematics of the $t\bar{t} + E_{T,miss}$ signature. They are therefore not a good framework for modeling this new physics signature.
On the other hand, the acceptances obtained with a merged sample of the leading-order process together with the radiation of an additional hard jet are in agreement with the NLO predictions. We also demonstrate  that the scale variation of the inclusive cross section, plus that of the acceptance, does not accurately reflect the uncertainty of the cross section after cuts, which is typically larger.  We show the importance of including higher-order QCD corrections when using kinematic distributions to determine the spin of the top-quark partner.

\end{abstract}

\maketitle

\thispagestyle{empty}
\end{titlepage}

\section{Introduction}

There is a strong theoretical expectation that the mechanism of electroweak symmetry breaking in Nature is more intricate 
than the single Higgs boson predicted by the Standard Model (SM).  One reason is that the large hierarchy between the Planck and electroweak scales is unstable in the SM.  The stabilization of this separation generically predicts the existence of new heavy partners of the top quark that cancel the quadratically divergent contribution of the SM top quark to the Higgs mass, thereby allowing the electroweak scale to be naturally small.  Another reason is that the relic abundance of the dark matter in the universe is naturally explained by a stable, neutral particle with a mass near the electroweak scale. Many extensions of the SM attempt to simultaneously solve both of these issues, and contain both a heavy new particle with the gauge quantum numbers of the SM top quark, and a new discrete symmetry which makes the lightest parity-odd particle a good dark matter candidate.  Examples of such models are the Minimal Supersymmetric Standard Model (MSSM)~\cite{Dimopoulos:1981zb} and the Littlest Higgs with T-parity~\cite{Cheng:2003ju}.  The MSSM contains a spin-0 stop quark and a spin-1/2 neutralino that fulfill the aforementioned roles, while the Littlest Higgs with T-parity contains a new spin-1/2 fermion and a parity-odd partner of the photon, which respectively serve as the top-quark partner and the dark matter candidate.  Although no hint of such states has yet been observed at the LHC, they remain the subject of intense theoretical and experimental interest.

A generic prediction of such theories is the QCD-initiated pair production of two top partners, followed subsequently by their decay into the SM top quarks plus the dark matter candidate.  This leads to the signature
\begin{equation}
 pp \to T \bar{T} \to t \bar{t} A_0 \bar{A_0}  \to t\bar{t} + E_{T,miss}.
\label{ttesig}
 \end{equation}
where $T$ generically denotes the top-quark partner and $A_0$ the dark-matter candidate. The $t\bar{t}$ pair then decays either semi-leptonically, or into a dilepton final state (we will not consider the fully-hadronic final state in this paper).  Such a process could be the dominant signature for supersymmetry in `natural SUSY' models that contain a light stop quark and a somewhat heavy gluino~\cite{Brust:2011tb}.  The signature of Eq.~(\ref{ttesig}) is also one of the simplified models suggested for presentation of LHC search results~\cite{Alves:2011wf}.  Top-quark plus missing energy signatures have been considered numerous times in the theoretical literature~\cite{Han:2008gy,Chen:2012uw}, and have been searched for experimentally~\cite{Aad:2011wc,Aad:2012uu,ATLAS-conf-2013-037,Aad:2012xqa,Chatrchyan:2013xna}.  The current limits exclude $T$ masses up to 600 GeV, depending on both the spin of the top partner and the mass of the dark matter particle. The proposed theoretical search strategies, and those utilized experimentally, all require an excess in the tail of an energy-related distribution, such as $E_{T,miss}$, the transverse mass of the lepton and missing $E_T$ if the top-quark pair decays semi-leptonically (denoted by $M_{T}$ in this manuscript), or the effective transverse mass $M_{T,eff} = E_{T,miss}+\sum_i E_{T,i}$, where $i$ runs over all observable particles and $E_{T,i} = \sqrt{m_i^2+\p_{T,i}^2}$.  It has been emphasized that variables such as $M_{T,eff}$ may also help distinguish the spin and other properties of the top partner~\cite{Chen:2012uw}.

In this manuscript we wish to improve upon the description of the $t\bar{t} + E_{T,miss}$ signal process, to assist in both the search for and eventual interpretation of the underlying model assuming discovery.  While the background processes can be probed in sideband regions, the signal-process description relies completely upon theory.  It is interesting to survey the current experimental analyses of this final state in order to understand how the signal process is modeled.  Although the background predictions are often data-driven, we list for comparison the simulation tools used to check and extrapolate the $t\bar{t}$ background.
\begin{itemize}

\item The analysis of Ref.~\cite{Aad:2011wc} searches for fermionic top partners in the dileptonic decay channel.  The signal is modeled using leading-order Madgraph~\cite{Alwall:2007st} matrix elements attached to a Pythia shower~\cite{Sjostrand:2006za}.  This prediction is then normalized to the inclusive approximate next-to-next-to-leading order (NNLO) QCD prediction from HATHOR~\cite{Aliev:2010zk}.  The $t\bar{t}$ background is modeled using MC@NLO~\cite{Frixione:2002ik}, normalized to an approximate NNLO prediction for the inclusive cross section~\cite{Moch:2008qy}.

\item The search of Ref.~\cite{Aad:2012uu} focuses on scalar top partners decaying to the semileptonic final state.  The signal is modeled using HERWIG++~\cite{Bahr:2008pv} and is normalized to an inclusive NLO result augmented with next-to-leading logarithmic (NLL) soft gluon resummation~\cite{Kramer:2012bx}.  The $t\bar{t}$ background is modeled using MC@NLO, and normalized to an approximate NNLO prediction.

\item The searches of Refs.~\cite{ATLAS-conf-2013-037} and~\cite{Aad:2012xqa} focus on scalar top partners in the semi-leptonic channel.  The signal is again modeled using HERWIG++ and normalized to an inclusive NLL+NLO prediction.  The
$t{\bar t}$ background is modeled using both MC@NLO and POWHEG~\cite{Nason:2004rx}, and is normalized to an approximate inclusive NNLO result.


\item The study of Ref.~\cite{Chatrchyan:2013xna} searches for scalar top partners in the semi-leptonic mode.  
For the signal events, the production of top-squark pairs is generated with MADGRAPH including up to two additional partons 
at the matrix element level.   This prediction is normalized to the inclusive NLL+NLO production cross section.  POWHEG is 
used to model the $t\bar{t}$ background, which is normalized to an approximate NNLO prediction for the inclusive cross section.

\end{itemize}  
We note that while the background kinematics is described using NLO QCD matched with a parton shower simulation, the signal is modeled using only leading-order matrix elements interfaced to a parton shower.  Although this level of simulation is sufficient for discovery of a dramatic new physics signature such as resonance production, it is not reliable when the signal is instead a subtle excess in the tail of a kinematic variable, as is expected to be the case for top-partner production.  In addition, the normalization to the inclusive NLO production cross section misses QCD effects in the decay chain.  Both of these deficiencies must be remedied to have the reliable signal predictions needed to enable discovery.

In a previous paper we began to improve upon modeling of the scalar top-partner signal process by deriving the NLO QCD correction to the $t\bar{t} + E_{T,miss}$ signature together with the semi-leptonic decay of the top-quark pair~\cite{Boughezal:2012zb}.  This was the first time that such a complex new physics signature was studied with exact NLO QCD corrections included consistently through the entire production and decay chain.  We found large, kinematic-dependent QCD corrections that differed significantly depending on the observable studied.  In this paper we extend this study in numerous ways.  We consider both scalar and fermionic top partners, and present predictions for LHC collisions at both 8 TeV and 14 TeV.  We also study both the semi-leptonic and dileptonic decay channels, and present a comprehensive analysis of higher order QCD effects on the relevant kinematic distributions that may aid in either exclusion,discovery or diagnosis.  In addition to comparing LO and NLO QCD predictions, we also study both leading-order plus parton shower simulations, and merged samples that contain the signal process with an additional hard jet.  We summarize below our main findings.
\begin{itemize}

\item In general, leading-order plus parton shower simulations do not correctly model the kinematics of the $t\bar{t} + E_{T,miss}$ signature.  Our study of Madgraph matrix elements for our signal process, with the `out-of-the-box' PYTHIA shower bundled with Madgraph, leads to acceptances which differ from the NLO predictions by nearly a factor of two.

\item The acceptances obtained with a merged sample of the leading-order process together with the radiation of an additional hard jet, interfaced to a parton shower, are in agreement with the NLO predictions.  The large differences between Pythia and NLO mentioned in the previous bullet are removed when a merged sample is considered.  Both NLO and the merged sample serves as accurate frameworks for the prediction of the $t\bar{t} + E_{T,miss}$ kinematics.

\item The tuned HERWIG++ simulations specifically used in the studies of Refs.~\cite{Aad:2012uu,ATLAS-conf-2013-037,Aad:2012xqa} produce acceptances in agreement with our NLO and merged predictions.  In future experimental studies, the acceptances and kinematic distributions obtained with leading-order plus parton-shower simulations should be compared to either the NLO prediction or a merged sample to ensure a correct description. 

\item The scale variation of the inclusive cross section does not accurately reflect the uncertainty of the cross section 
after experimental cuts are imposed, especially in the semi-leptonic channel.  The latter uncertainty is typically larger, and should be used as the theoretical systematic error in experimental analyses.

\end{itemize}

Our manuscript is organized as follows.  We formulate our study, and present details of the calculation in Section~\ref{sec:setup}.  We discuss our numerical results in detail in Section~\ref{sec:numerics}.  Finally, we conclude in Section~\ref{sec:conc}.

\section{Setup}
\label{sec:setup}

We consider both scalar and fermionic top partners in our study.  The signature we focus on is
\begin{equation}
 pp \to T \bar{T} \to t \bar{t} A_0 \bar{A_0}  \to t\bar{t} + E_{T,miss},
 \end{equation}
where $T$ generically denotes the top-quark partner, and $A_0$ the dark-matter candidate.  The decay of the top partner in both cases is $T \to A_0 t$.  In the case of scalar top partners, $A_0$ is a Majorana fermion, while in the fermonic case it is a massive spin-one vector particle.  Our results for scalar top partner cover the Minimal Supersymmetric Standard Model~\cite{Dimopoulos:1981zb} in the heavy-gluino limit for any choice of stop quark mixing, while our fermionic model covers the Littlest Higgs with T-parity~\cite{Cheng:2003ju}; we discuss in more detail later in this section exactly how our simplified model reproduces the MSSM.  For the decay of the top quarks, we study observables for both the semi-leptonic and dileptonic final states.  We study this process at both leading-order and through next-to-leading order in the QCD coupling constant, with NLO effects included throughout the entire decay chain.  We also compare these fixed-order results with a parton shower simulation matched to exact leading-order matrix elements, as used in experimental searches for this signature, and also to a merged sample containing an additional hard jet.

Since we present a larger number of numerical results in this paper, we summarize below the various parameter choices considered.
\begin{itemize}

\item We study scalar top-partner production in the semi-leptonic decay mode at an 8 TeV LHC.  Our analysis follows the ATLAS searches discussed in Refs.~\cite{ATLAS-conf-2013-037,Aad:2012xqa}.  We study two parameter points: $m_T= 600$ GeV, $m_{A_0}=50$ GeV, which is slightly above the current exclusions limit; $m_T= 225$ GeV, $m_{A_0}=25$ GeV, which corresponds to a compressed spectrum and is not currently excluded.  

\item We study fermionic top-partner production at an 8 TeV LHC in the semi-leptonic decay mode for the parameter choice $m_T= 600$ GeV, $m_{A_0}=50$ GeV, in order to compare the QCD corrections with those affecting the scalar partner.

\item We study both fermionic and scalar partner production at a 14 TeV LHC in the di-leptonic decay mode for the parameter choice $m_T= 600$ GeV, $m_{A_0}=50$ GeV.  We also perform a detailed comparison of scalar and fermion distributions for several parameter choices, to study the effect of QCD corrections on the discrimination between the two spin possibilities.

\end{itemize}

\subsection{Calculational framework}

In order to perform our analysis we must calculate the fully-differential cross sections through NLO in QCD.  We briefly describe the techniques used to obtain our results.  Although this was discussed in our 
previous work~\cite{Boughezal:2012zb}, we repeat the discussion here for completeness.  We calculate the NLO QCD corrections to the processes 
$pp \rightarrow T \bar T \rightarrow b \bar b l \nu j j A_0 \bar A_0$ and $pp \rightarrow T \bar T \rightarrow b \bar b l \bar l \nu  \bar \nu A_0 \bar A_0$
by extending the framework of Ref.~\cite{Melnikov:2009dn} for top-quark pair production.
We assume the production of a scalar or fermonic $T \bar T$ pair which is followed by 
consecutive on-shell decays of $T \rightarrow t A_0$, $t \rightarrow b W$ and $W \rightarrow l\nu / jj$.
We assume that the top partner decays $100\%$ of the time through the process $T \rightarrow t A_0$.
We neglect contributions that are parametrically suppressed by 
$\mathcal{O}(\Gamma_T / m_T)$, $\mathcal{O}(\Gamma_t / m_t)$ and $\mathcal{O}(\Gamma_W / m_W)$, in each of the decay stages respectively.  At leading-order in the perturbative QCD expansion, both $gg$ and $q\bar{q}$ partonic channels contribute.  The $qg$ initial state begins to contribute at NLO.

This sequential framework is then systematically promoted to NLO accuracy by calculating QCD corrections 
to the production and decay processes, including all spin correlations in the narrow-width approximation.
If desired, we can systematically improve our approximation by allowing off-shell top quarks.
We numerically calculate virtual corrections for the production process
via $D$-dimensional generalized unitarity methods~\cite{Giele:2008ve}.  For the case of the scalar top partner, 
we extend these techniques by deriving new tree-level recursion currents involving scalars, quarks and gluons. For the 
fermonic top partner, no new currents are required.  Real corrections to $T\bar T$ pair production do not exhibit final-state collinear singularities and 
soft singularities are spin-independent, allowing us to reuse previous results for top quarks~\cite{Melnikov:2009dn}.
QCD corrections to the decay $T\rightarrow A_0 t$ are derived analytically using a traditional Feynman-diagrammatic approach.  We can make use of existing results for top quarks to treat singularities in the real-emission decay process.
We subtract the soft singularity in $T \rightarrow A_0 t g $ with the dipoles of Ref.~\cite{Campbell:2012uf} 
which were developed for the decay $t\rightarrow W b$ retaining a finite $b$-quark mass.
QCD corrections to the remaining stages in the decay chain, $t \rightarrow b W$ and $W \rightarrow jj$, are taken
from previous results for top-pair production.

\subsection{Checks of the result}
 
We comment briefly here on the checks we performed to ensure the correctness of our results.  First, we confirmed numerically that $1/\varepsilon$-poles in dimensional regularization, where $\varepsilon = (4-d)/2$,
cancel between virtual and real corrections in the production as well as in the decay matrix elements.
To check the finite parts, the virtual corrections to the process $q\bar q \rightarrow T \bar T$, where $T$ denotes either spin possibility, were calculated with an
independent Feynman diagrammatic calculation for stable top partners, and complete agreement was found.
The virtual corrections to the top-partner decay processes were also cross-checked by a second independent
calculation.  The implementation of the real corrections was checked for independence on the cut-off parameter $\alpha$ 
that controls the resolved phase space of the dipole subtraction terms.  To further check the implementation of the decay stages, we tested factorization properties between production and decay matrix elements.
This is achieved by removing all acceptance cuts on final state-particles and integrating over the full phase space. 
The result is compared to a separate evaluation of the product of total cross section for stable $T \bar{T}$ pairs times 
their branching fraction.  We find that the required identities are fulfilled within the numerical precision.

\begin{figure}[h!]
\centerline{
\includegraphics[width=12.2cm]{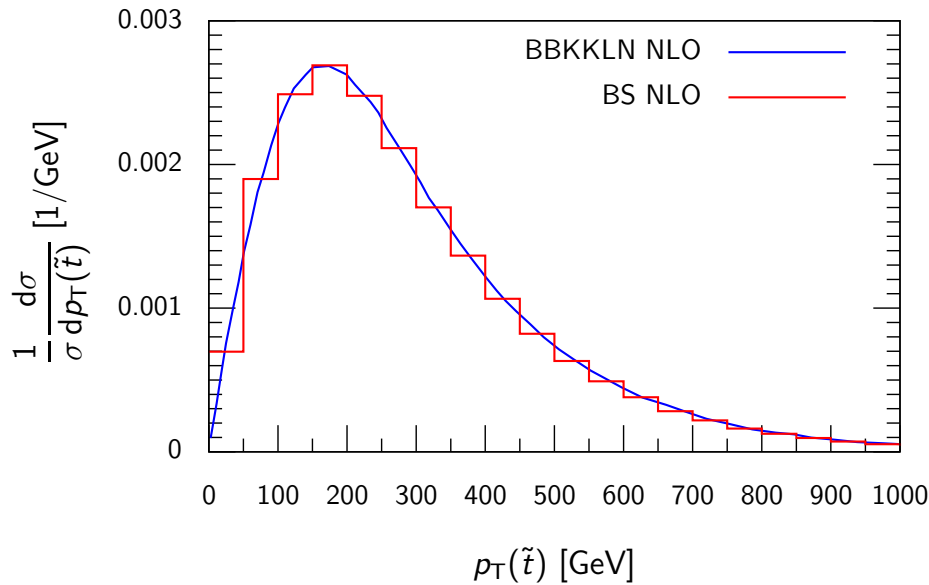}
}
\caption{A comparison of the scalar-top transverse momentum spectrum for $m_{\tilde t}$=500"~GeV at a 14 TeV LHC between our calculation (BS) and the one from Ref.~\cite{Beenakker:2010nq} (BBKKLN). }
\label{fig:check-PT-BBKKLNvsBS}
\end{figure}

We additionally compared our results against those available in the literature where possible.  For scalar top-partners, we 
compared the inclusive cross section to the results of Ref.~\cite{Beenakker:1997ut} as implemented in \verb+Prospino 2.1+ \cite{Beenakker:1996ed} in the heavy-gluino limit.  Agreement between the hadronic cross sections at the 0.1\% level was found.  We also compared our result for the scalar-top transverse momentum spectrum with the one from Ref.~\cite{Beenakker:2010nq}, and found complete agreement (see Fig.~\ref{fig:check-PT-BBKKLNvsBS}).  For fermionic top partners, we compared against the NLO inclusive hadronic cross section of  \verb+Hathor+ \cite{Aliev:2010zk}, and found agreement at the 0.2\% level.  All of these checks together give us confidence to proceed with a detailed numerical study of top-partner production at the LHC. 

Before continuing, we comment in more detail on exactly how our result reproduces stop-pair production in the MSSM.  At NLO, the stop production cross section depends on three additional parameters besides the stop mass: the gluino mass, the stop mixing angle, and the light-flavor squark masses. The dependence of the cross section on these additional parameters was found to be at most 2\% in several example SUSY models in Ref.~\cite{Beenakker:2010nq}. We also confirm using Prospino that this production channel receives negligible gluino contributions from heavy virtual gluinos once its mass exceeds one TeV. This choice is well motivated since
lighter gluinos are already experimentally excluded for a wide variety of models. The decay of a stop quark depends on additional electroweak parameters as well as the stop mixing angle and the mixing matrix of the neutral gauge eigenstates. However, if we assume a 100 percent branching fraction of the stop quark into a neutralino and a top quark, only the relative strength of left- and right-handed coupling remains.

\section{Numerical results}
\label{sec:numerics}

In this section we present and discuss in detail our numerical results.  We show predictions for top-partner production using four different types of simulation: leading order in QCD; next-to-leading order in QCD; a leading-order top-partner production sample with parton-showering included; a sample of top-partner production with one additional hard jet merged with a parton shower by the MLM procedure.  For the latter two we use Madgraph~\cite{Alwall:2011uj} to generate the tree-level matrix elements and Pythia~\cite{Sjostrand:2006za} to perform the showering.  We use Pythia version 6.426 as provided in the default Madgraph distribution.  All the shown uncertainties are 
obtained by varying the renormalization and factorization scales by a factor of two around the top partner mass. While the production of the top partners is completely determined by QCD gauge invariance, the top-partner decay depends on the top-partner coupling to $A_0$.  Our choice of the left- and right-handed couplings for the scalar top partner to the top and $A_0$ follows the ATLAS parameters choice whereas for the fermionic top-partner our choice is as follows:
\begin{equation}
g_R =c_R m_t/v; g_L =c_L m_t/v,
\end{equation}
with $c_R =3/10$, $c_L =1/10$, $m_t =172$ GeV and $v=246$ GeV. We note that this choice does not have a strong theoretical motivation.  It is meant to illustrate the impact of higher-order effects, and to make connection to our previous work~\cite{Boughezal:2012zb}.

\subsection{Top partners at an 8 TeV LHC}

\begin{figure}[ht!]
\centerline{
\includegraphics[width=8.2cm]{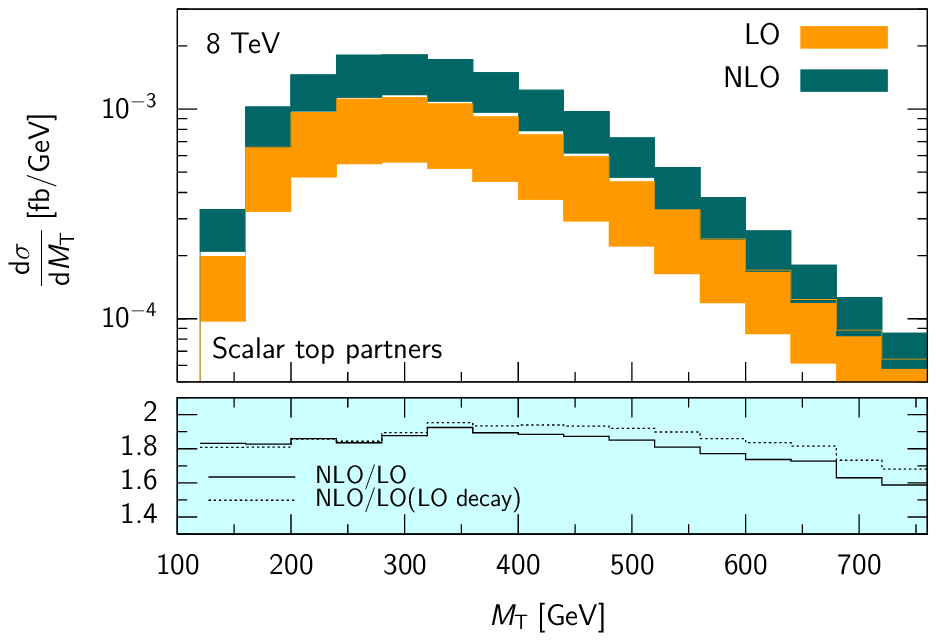}
\includegraphics[width=8.2cm]{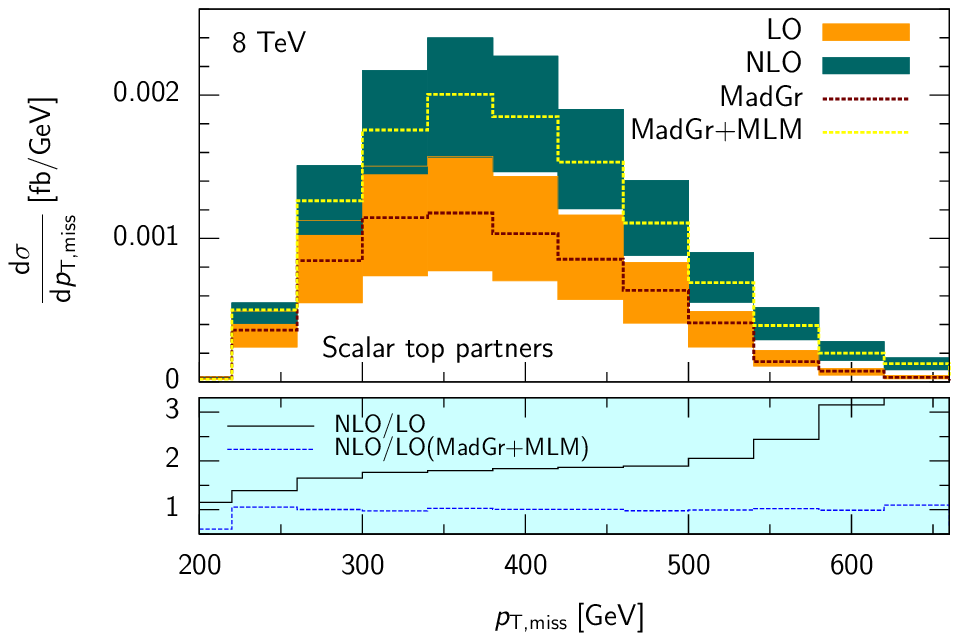}
}
\caption{The transverse mass (left) and missing transverse momentum (right) at LO and at NLO for scalar top partners with $(m_T,m_{A_0}) = (600\text{GeV},50\text{GeV})$ and
a semi-leptonic decay of the top.  The upper panel shows the distributions, while the lower panel shows the $K$-factors, defined as the ratio of NLO over LO.  For the $p_{T,miss}$ distribution, the Madgraph and merged-sample (labeled MadGr+MLM) results are also shown.}
\label{fig:stops.LHC8.600-50.semi}
\end{figure}

\begin{figure}[h!]
\centerline{
\includegraphics[width=10.0cm]{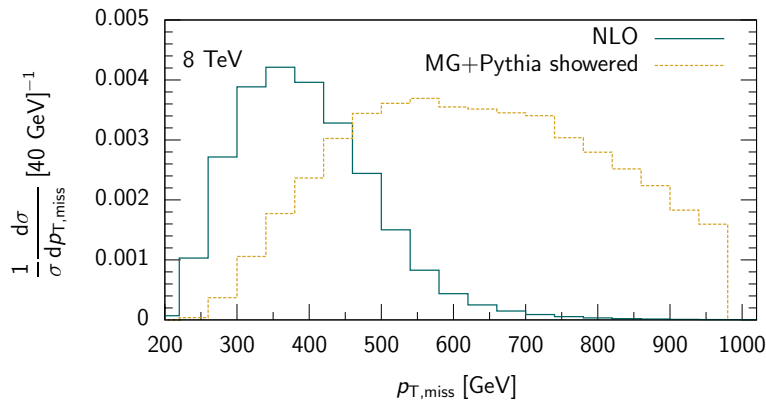}
}
\caption{The missing transverse momentum distributions obtained using our NLO result, and from the `out-of-the-box' Pythia settings bundled with Madgraph.}
\label{fig:stops-LHC8-600-50-semi-pTmiss_shape2}
\end{figure}

\begin{figure}[h!]
\centerline{
\includegraphics[width=10.0cm]{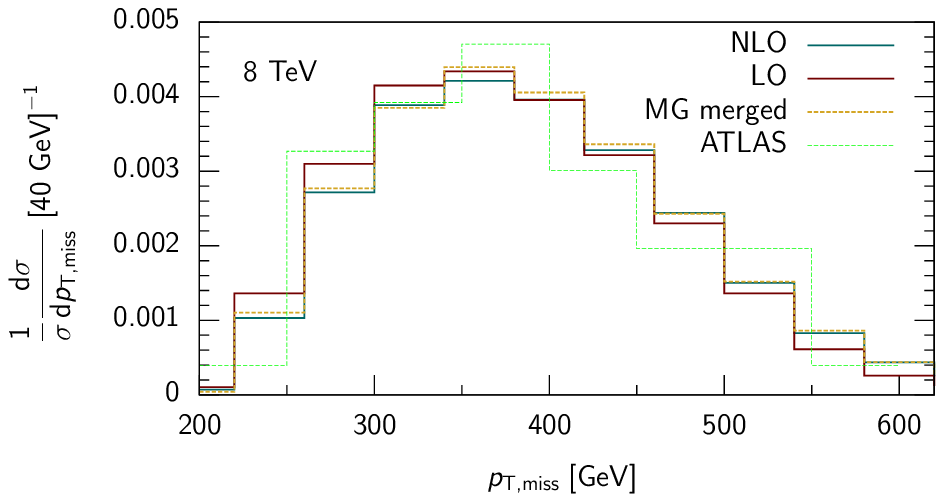}
}
\caption{A comparison of the $p_{T,miss}$ distribution used in the ATLAS study of Ref.~\cite{Aad:2012xqa} with those obtained using LO QCD, NLO QCD, and a merged sample.}
\label{fig:stops-LHC8-600-50-semi-pTmiss_shape1}
\end{figure}

We begin by presenting results for scalar top-partner production at an 8 TeV LHC, assuming the parameter values $m_T= 600$ GeV and $m_{A_0}=50$ GeV.  These choices are near the current exclusion limits set by ATLAS~\cite{ATLAS-conf-2013-037}.  We note that the variables used to discriminate signal from background in this analysis where the missing transverse momentum $p_{T,miss}$ and the transverse mass $M_T$, defined as \[M_{T} = 2 p_{T\,l} \, E_{T,miss} \, ( 1- \cos(\Delta \phi) ).\]  We therefore focus on them in our study.  The effective transverse mass $M_{T,eff}$, defined as the scalar sum of $p_{T,miss}$ and the transverse momenta of all final-state jets, has been suggested as a useful diagnostic to determine the top-partner spin~\cite{Chen:2012uw}, and we include it also in our study.  We  impose the following acceptance cuts, which are similar to those used in the ATLAS analysis:
%

\begin{eqnarray}
    \Delta R_j&=& 0.4,\;\; p_{Tj } >  30 \,\text{GeV}, \;\; |y_j| <  2.5,\nonumber \\ 
    p_{Tl} &>&  25 \,\text{GeV}, \;\; |y_{l}| <  2.4, \;\;p_{T,miss} >  150 \,\text{GeV}\nonumber \\ 
    M_{T}  &>&   140 \,\text{GeV}, \;\; p_{T,miss}/\sqrt{H_T}>13  \sqrt{GeV},
    \label{cuts}
\end{eqnarray}  
\noindent
where $H_T$ is the scalar sum of the momenta of all final-state jets. Shown in Fig.~\ref{fig:stops.LHC8.600-50.semi} are the $M_T$ and $p_{T,miss}$ distributions for scalar top-partner production using various different simulation tools. For $M_T$, results at LO and NLO in fixed-order QCD are shown. The shape change when going from LO to NLO amounts to up to 20\%, and leads to an overall normalization shift of approximately 80\% over most of the studied range.  We have observed no difference in the $M_T$ shape when comparing the fixed-order results with those from the matrix-element plus parton shower simulation.  We also show in this plot the effect of calculating the $T$ decay to only leading order, and find that the shape changes by up to 10\% due to QCD corrections in the decay.
The situation is different for the $p_{T,miss}$ distribution.  The $K$-factor grows large with increasing $p_{T,miss}$, increasing from 1.2 at $p_{T,miss}=150$ GeV to over 2 at $p_{T,miss}=500$ GeV.  This changes the acceptance which enters the extrapolation of the fiducial cross section to the inclusive one bounded in the experimental analysis.  We also show on this plot the distributions obtained using a leading-order Madgraph analysis, and with a merged sample also containing an additional hard jet radiated along with the top-partner pair.  The agreement of the Madgraph curve with our leading-order results provides additional validation of our results.  The agreement of the merged sample with the NLO curve suggests that it also provides a good framework for calculating the acceptance.  We compare in Table~\ref{table:stopacc} the acceptances calculated using four different tools: LO QCD, NLO QCD, LO with Pythia showering included, and a leading-order merged sample with zero and one additional hard jet.  There is a large shift of over 40\% in the acceptance when going from LO to NLO.  The LO scale dependence vanishes, and is not a good estimate of the higher-order corrections.  The NLO acceptance agrees well with the value obtained using the matched sample, indicating that the shape difference when going from LO to NLO comes from the emission of an additional hard jet.  Both the NLO and merged sample serve as a good framework for predicting the shapes of the signal distribution.  Of course, the merged sample is based on a leading-order calculation and comes with a significantly larger normalization uncertainty. Interestingly, the `out-of-the-box' Pythia simulation that is bundled with Madgraph gives a much larger acceptance than the NLO or the merged sample, due to a much harder $p_{T,miss}$ spectrum produced by the shower.  This spectrum is shown in Fig.~\ref{fig:stops-LHC8-600-50-semi-pTmiss_shape2}, along with that obtained at NLO.  This illustrates the danger of using a pure parton shower result in analyses.  We note, however, that the 
tuned Herwig simulation used in Refs.~\cite{ATLAS-conf-2013-037,Aad:2012xqa} does accurately reproduce the NLO distribution shapes.  This is shown in Fig.~\ref{fig:stops-LHC8-600-50-semi-pTmiss_shape1}, where the distribution shapes from ATLAS are compared to those from the various tools considered in our study.  This indicates that the acceptance predictions used in these studies are close to the correct NLO value.

\begin{table}[htbp]
\begin{center}
\begin{tabular}{|c || c | c | c | c |}
\hline
  & LO & NLO & MG+Pythia & MG+PS merged \\ \hline \hline
acceptance & $0.19^{+0}_{-0}$ & $0.27^{+0.3}_{-0.2}$ & 0.46 & 0.27 \\ \hline
\end{tabular}
\end{center}
\caption{Acceptances for a scalar top partner and a semi-leptonic decay of the top assuming the experimental cuts shown in Eq.~(\ref{cuts}) with four different simulations: LO, NLO, LO with Pythia showering included, and a leading-order merged sample with zero and one additional hard jet.}
\label{table:stopacc}
\end{table}

\begin{figure}[h!]
\centerline{
\includegraphics[width=10cm]{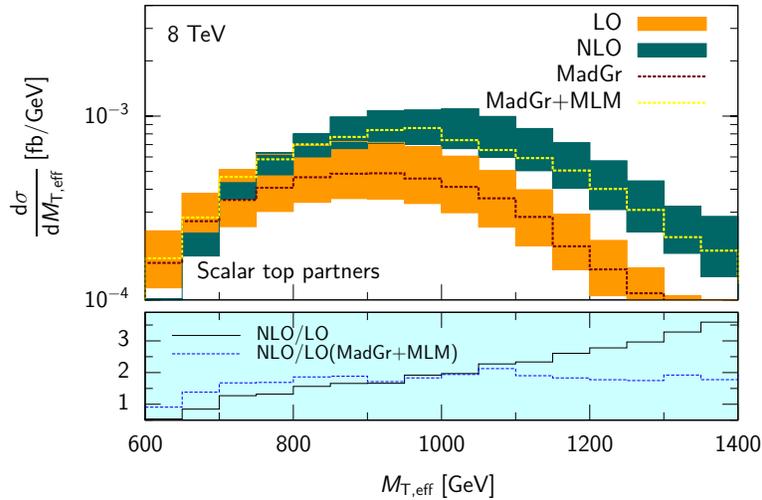}}
\caption{The effective transverse mass at LO and at NLO both with and without corrections throughout the entire decay chain for 
the scalar-top partner with $(m_T,m_{A_0}) = (600\text{GeV},50\text{GeV})$.  Also shown is the $M_{T,eff}$ distribution for the LO-Madgraph and merged Madgraph+Pythia sample.  The upper panel shows the distributions, while the lower panel shows the $K$-factors.}
\label{fig:stops.LHC8.600-50.semi.MTeff}
\end{figure}

We show in Fig.~\ref{fig:stops.LHC8.600-50.semi.MTeff} the $M_{T,eff}$ distribution.  In our previous work we found a large $K$-factor for this process, which reached over three~\cite{Boughezal:2012zb}.  We identified the reason for this result; to obtain the required four jets to pass the analysis cuts, all four partons in the LO final state must be well-separated.  At high $M_{T,eff}$, the top partners in the final state become boosted, and the partons fall inside the same jet cone.  This is alleviated at NLO, when an additional parton is radiated into the final state.  The merged sample does not suffer from this problem, and the $K$-factor when going from LO to the merged calculation remains relatively flat as a function of $M_{T,eff}$.  However, it does not correctly reproduce the normalization of the NLO result, which is approximately 80\% higher.

\begin{figure}[h!]
\centerline{
\includegraphics[width=8.2cm]{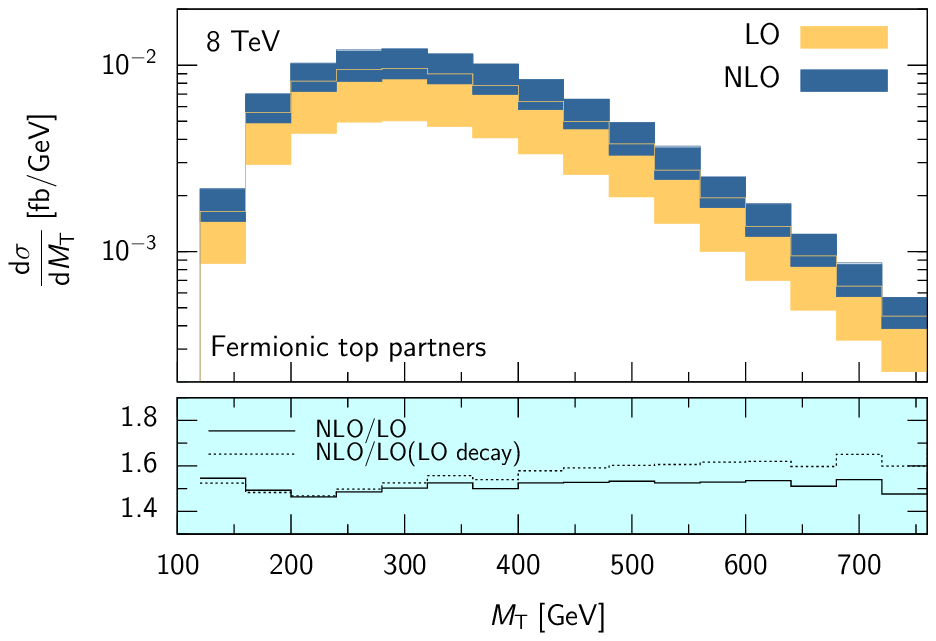}
\includegraphics[width=8.2cm]{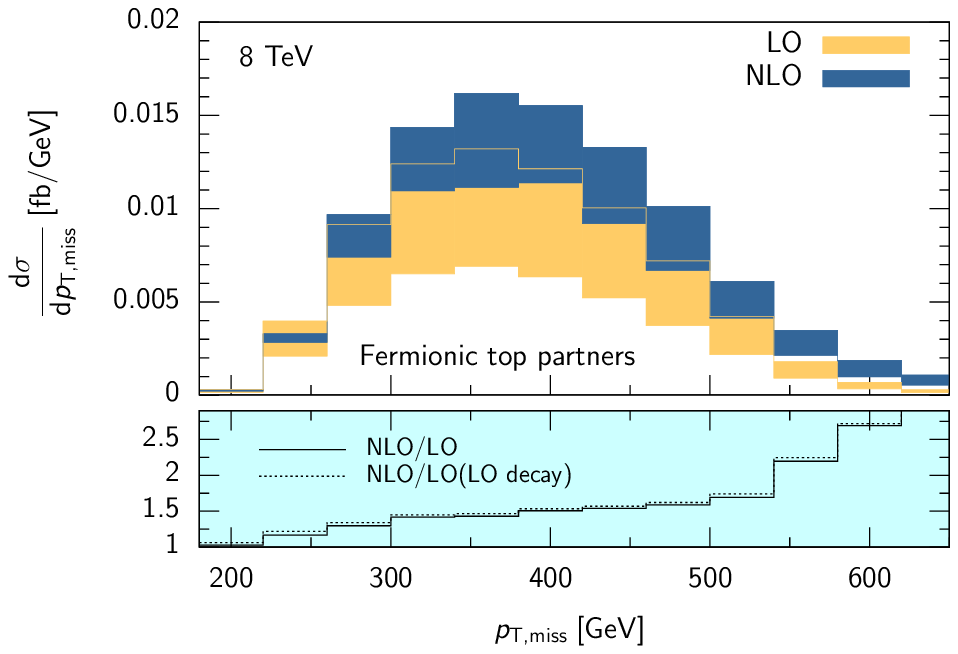}
}
\caption{The transverse mass (left) and missing transverse momentum (right) at LO and at NLO both with and without corrections throughout the entire decay chain for fermionic top partners with $(m_T,m_{A_0}) = (600\text{GeV},50\text{GeV})$.  The upper panel shows the distributions, while the lower panel shows the $K$-factors, defined as the ratio of NLO over LO.}
\label{fig:tprimes.LHC8.600-50.semi}
\end{figure}

We now compare these result to those of fermionic top partner production, also with $m_T= 600$ GeV and $m_{A_0}=50$ GeV.  We show in Fig.~\ref{fig:tprimes.LHC8.600-50.semi} the $M_T$ and $p_{T,miss}$ distributions at LO and NLO in perturbative QCD.  The $K$-factor for the $M_T$ distribution is flat, like in the scalar partner case, with an increase of roughly 40\% over the entire spectrum.  Neglecting QCD corrections in the decay would induce errors at the 5-10\% level in the tail of the distribution.  For the $p_{T,miss}$ distribution, the NLO QCD correction increases the rate by an amount starting from zero at $p_{T,miss}=150$ GeV, and increasing to over 50\% at $p_{T,miss}=500$ GeV.  It must be included for an accurate prediction of the spectrum.  We show the cross sections before and after cuts, together with the acceptances, according to LO and NLO QCD in Table~\ref{table:fermacc}.  There are several points to make about these results.  We first note that the scale dependences of the inclusive cross section, and the cross section after cuts, show different behavior.  For simplicity we symmetrize the upper and lower scale variations in this discussion.  The scale dependence of the inclusive cross section decreases from $\pm 33\%$ to $\pm 14\%$ when going from LO to NLO.  For the cross section after cuts, it changes from $\pm 34\%$ to $\pm 18\%$.  The scale dependence of the inclusive cross section cannot be used to estimate the theoretical uncertainty for the signal cross section that enters the experimental analysis, as the cross section after cuts exhibits a larger scale uncertainty.  This behavior appears generic; we will see it again in the next section when studying a compressed spectrum.  Again, as for the scalar top partners, the scale variation of the acceptance does not accurately reflect this shift that occurs when going from LO to NLO.  We note that the acceptance is nearly identical to that of the scalar top partner shown in Table~\ref{table:stopacc}.  While this suggests the current experimental bounds on scalar top-partner production can be used to also constrain fermion top partners, it also indicates that other variables besides $M_T$ and $p_{T,miss}$ will be needed to discriminate between spin possibilities if a future discovery is made.  We revisit this point in a later section.

\begin{table}[htbp]
\begin{center}
\begin{tabular}{| c || c | c | c |}
\hline
 & $\sigma_{total}$ (fb)& $\sigma_{cut}$ (fb) & acceptance \\ \hline \hline
LO & $10.79^{+4.35}_{-2.86}$ & $2.15^{+0.87}_{-0.57}$ & $0.20^{+0}_{-0}$  \\ \hline
NLO & $12.1^{+1.3}_{-2.1}$ & $3.23^{+0.62}_{-0.55}$ & $0.27^{+0.02}_{-0}$ \\ \hline
\end{tabular}
\end{center}
\caption{Inclusive cross section, the cross section after the experimental cuts shown in Eq.~(\ref{cuts}), and the acceptance for a fermonic top partner with a semi-leptonic decay of the top-quark, at LO and NLO in QCD.  The uncertainties coming from QCD scale variation are shown for each quantity.}
\label{table:fermacc}
\end{table}

\subsection{A compressed spectrum at an 8 TeV LHC}

\begin{figure}[h!]
\centerline{
\includegraphics[width=8.2cm]{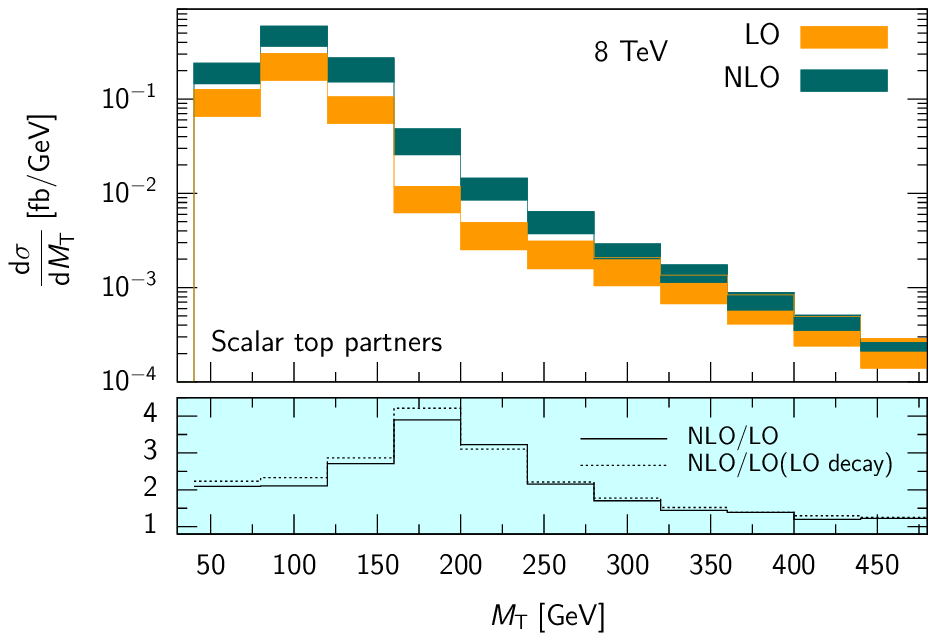}
\includegraphics[width=8.2cm]{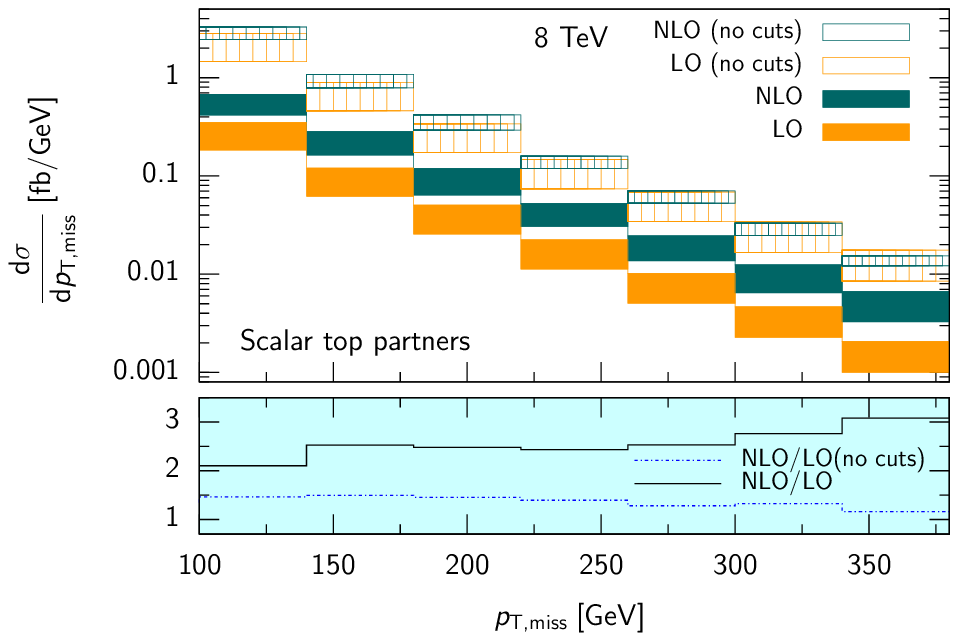}
}
\caption{The transverse mass (left) and missing transverse momentum (right) at LO and at NLO both with and without corrections throughout the entire decay chain for scalar top partners with $(m_T,m_{A_0}) = (225\text{GeV},25\text{GeV})$.  For the $p_{T,miss}$ result we show the results at LO and NLO both before and after including the cuts of Eq.~(\ref{cuts2}).  The upper panel shows the distributions, while the lower panel shows the $K$-factors.}
\label{fig:stops.LHC8.225-25.semi}
\end{figure}

We now study a scalar top-partner with the following parameters: $m_T= 225$ GeV and $m_{A_0}=25$ GeV.  Since the top quark has a mass of approximately 175 GeV, there is little kinetic energy released in the decay $T \to t A_0$.  This is an example of a ``compressed spectrum" in which light supersymmetric particles may still evade LHC search constraints~\cite{LeCompte:2011cn}.  We follow the ATLAS analysis of Ref.~\cite{ATLAS-conf-2013-037} and utilize the following experimental cuts in presenting our results for the compressed spectrum:

%
\begin{eqnarray}
    \Delta R_j&=& 0.4,\;\; p_{Tj } >  30 \,\text{GeV}, \;\; |y_j| < 2.5, \nonumber \\ 
    p_{Tl} &>&  25 \,\text{GeV}, \;\; |y_{l}| <  2.4,\;\;p_{T,miss} >  150 \,\text{GeV}  \nonumber \\ 
    M_{T} & >&   140 \,\text{GeV}, \;\; 
    p_{T,miss}/\sqrt{H_T}>5 \sqrt{GeV}
    \label{cuts2}
\end{eqnarray}  
where $H_T$ is the scalar sum of the momenta of all final-state jets.  We show in Fig.~\ref{fig:stops.LHC8.225-25.semi} the $M_T$ and $p_{T,miss}$ distributions.  For the $p_{T,miss}$ result we show the results at LO and NLO both before and after including the cuts of Eq.~(\ref{cuts2}).  Before the cuts are imposed, the $K$-factor takes on a constant value of approximately 1.5 over the entire spectrum.  After cuts are imposed, it changes to roughly two at the lower boundary, and rises to three at $p_{T,miss}=350$ GeV.  The corrections to the $M_T$ distribution also show a strong kinematic dependence after the selection cuts are imposed.  We note that the behavior of the $M_{T,eff}$ distribution at LO, NLO and with the merged sample is similar to that found in the previous section.

The cross sections before and after cuts, together with the acceptances, according to LO and NLO QCD are shown in Table~\ref{table:compacc}. Like for the fermionic top partner discussed in the previous section, the selection cuts significantly change the theoretical uncertainty as estimated by scale variation.  The scale variation of the inclusive cross section decreases from $\pm 34\%$ at LO to $\pm 14\%$ at NLO.  It only decreases from $\pm 34\%$ at LO to $\pm 27\%$ at NLO after cuts are imposed.  The estimated theoretical error is twice as large as would be estimated by using the inclusive cross section.  The theoretical error derived from scale variation of the inclusive cross section and the acceptance does not accurately reflect the real uncertainty present in the fiducial cross section to which the experimental analysis is sensitive.

\begin{table}[htbp]
\begin{center}
\begin{tabular}{| c || c | c | c |}
\hline
 & $\sigma_{total}$ (fb)& $\sigma_{cut}$ (fb) & acceptance \\ \hline \hline
LO & $530^{+216}_{-141}$ & $15.88^{+6.57}_{-4.27}$ & $0.03^{+0}_{-0}$  \\ \hline
NLO & $722^{+95}_{-99}$ & $36.0^{+11.3}_{-7.9}$ & $0.05^{+0.01}_{-0}$ \\ \hline
\end{tabular}
\end{center}
\caption{Inclusive cross section, the cross section after the experimental cuts shown in Eq.~(\ref{cuts}), and the acceptance for a compressed-spectrum scalar top partner, at LO and NLO in QCD.  The uncertainties coming from QCD scale variation are shown for each quantity.}
\label{table:compacc}
\end{table}

\begin{figure}[h!]
\centerline{
\includegraphics[width=12.2cm]{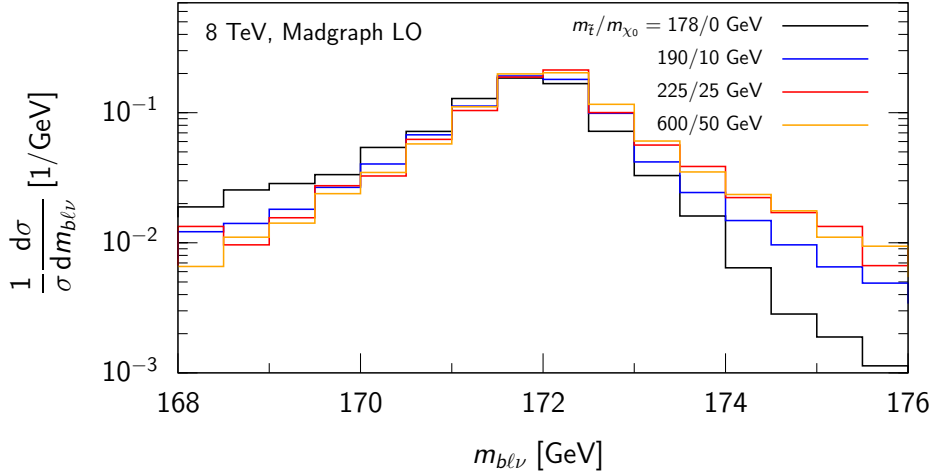}
}
\caption{Estimate of top quark off-shell effects in stop quark decays for several choices of  $(m_{\tilde t},m_{A_0})$ configurations. Shown is the fraction of events in each bin of the invariant mass of the leptonically decaying top-quark. The Breit-Wigner distribution around the top mass remains undistorted for the configuration (225,25) GeV indicating the validity of the narrow-width approximation for the compressed-spectrum mass choice.}
\label{fig:offshellness}
\end{figure}

We finally study the impact of top-quark off-shell effects for light stop quarks decays for several $(m_{\tilde t},m_{A_0})$  configurations.
As studied in Ref.~\cite{Kilic:2012kw}, the subsequent top-quark decay can proceed through off-shell top quarks and lead to significant distortions of the kinematic distributions. We have checked that for our mass choice (225/25) GeV the Breit-Wigner line shape of the top quark peak remains intact (see  Fig.~\ref{fig:offshellness}) and the narrow-width approximation is still applicable.

\subsection{The dileptonic mode at a 14 TeV LHC}

\begin{figure}[h!]
\centerline{
\includegraphics[width=8.2cm]{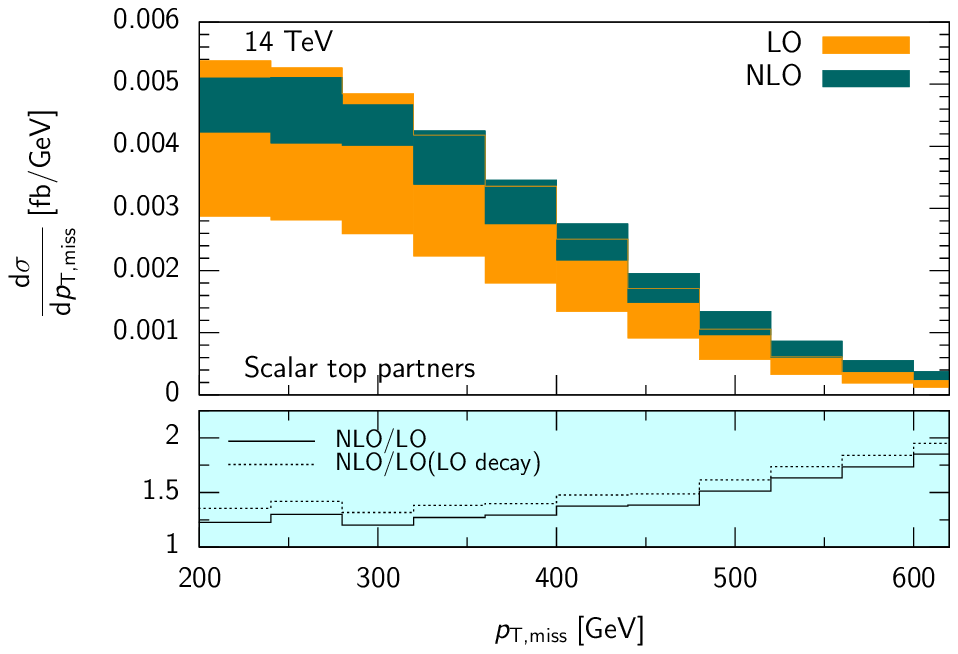}
\includegraphics[width=8.2cm]{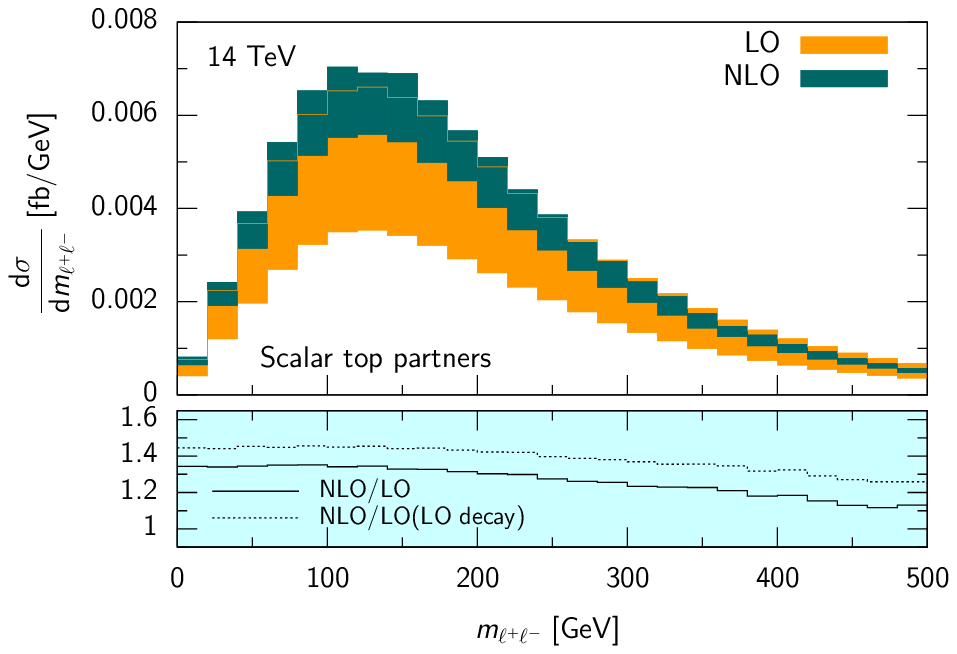}}
\caption{Distributions for scalar top partners with $(m_T,m_{A_0}) = (600\text{GeV},50\text{GeV})$ at a 14 TeV LHC.  The distributions shown are: the missing transverse momentum and the dileptonic invariant mass.}
\label{fig:stops.LHC14.600-50.semi1}
\end{figure}

\begin{figure}[h!]
\centerline{
\includegraphics[width=8.2cm]{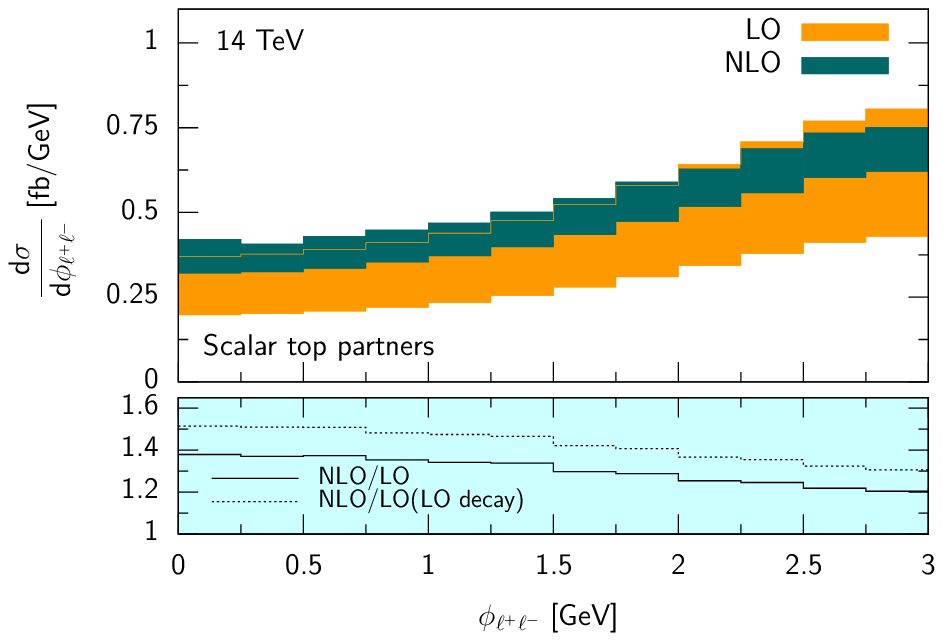}
\includegraphics[width=8.2cm]{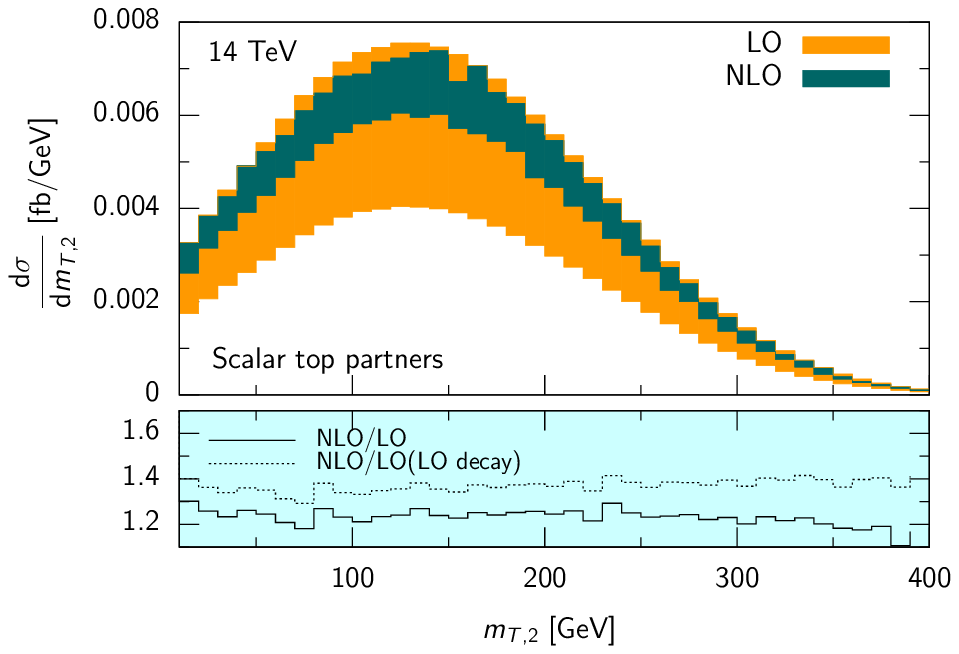}
}
\caption{Distributions for scalar top partners with $(m_T,m_{A_0}) = (600\text{GeV},50\text{GeV})$ at a 14 TeV LHC.  The distributions shown are the transverse-plane opening angle between the two leptons and $m_{T,2}$. }
\label{fig:stops.LHC14.600-50.semi2}
\end{figure}

We now consider searches at a future 14 TeV run of the LHC.  We focus on the dileptonic final state, in order to also illustrate the effect of higher-order QCD on this channel.  This final state is experimentally cleaner, due to the smaller number of final-state jets.  The dominant background now becomes $t\bar{t}$ production in the dileptonic decay mode.  Due to the presence of neutrinos in the decay of both top quarks, the transverse mass no longer effectively discriminates signal from background, and different variables must be used instead.  In addition to the previously considered $p_{T,miss}$, we also present results for three other variables: $\phi_{l^+l^-}$, the angle between the two leptons in the transverse plane; $m_{l^+l^-}$, the dilepton invariant mass; $m_{T,2}$, as defined in Ref.~\cite{Lester:1999tx}.  We show in Figs.~\ref{fig:stops.LHC14.600-50.semi1} and~\ref{fig:stops.LHC14.600-50.semi2} the results for all four variables at both LO and NLO in QCD.  The $K$-factor for $p_{T,miss}$ exhibits a strong kinematic dependence, like in the semi-leptonic mode at the 8 TeV LHC.  However, the other variables receive a fairly flat QCD correction, with a variation of 20\% or less over the entire kinematic range.  We note that neglecting the QCD corrections in the decay induces roughly 10\% errors in the theoretical predictions.

\begin{table}[htbp]
\begin{center}
\begin{tabular}{| c || c | c | c || c | c |}
\hline
 & $\sigma_{total}$ (fb)& $\sigma_{cut}$ (fb) & acceptance & $\sigma_{cuts+m_{T,2} \, cut}$ (fb) & acceptance \\ \hline \hline
LO &  $2.06^{+0.80}_{-0.53}$ & $1.25^{+0.49}_{-0.32}$ & $0.61^{+0.00}_{-0.00}$ & $0.71^{+0.27}_{-0.19}$ & $0.35^{+0.00}_{-0.01}$ \\ \hline
NLO & $2.72^{+0.28}_{-0.34}$ & $1.61^{+0.15}_{-0.19}$ & $0.59^{+0.01}_{-0.00}$ & $0.88^{+0.07}_{-0.10}$ & $0.32^{+0.00}_{-0.01}$  \\ \hline
\end{tabular}
\end{center}
\caption{Inclusive cross section, the cross section after the experimental cuts shown in Eq.~(\ref{cuts3}), and the acceptance for a scalar top partner, at LO and NLO in QCD, for a 14 TeV LHC. The right part of the table shows the cross section and acceptance after also including a cut on $m_{T,2}$}
\label{table:14TeVstop}
\end{table}

We show in Table~\ref{table:14TeVstop} the cross sections before and after the following cuts:

\begin{eqnarray}
    \Delta R_j&=& 0.4,\;\;  p_{Tj } >  50 \,\text{GeV}, \nonumber \\  
    |y_j| &< & 2.5,\;\; p_{Tl} >  20 \,\text{GeV}, \;\; |y_{l}| <  2.5, \nonumber \\ 
    p_{T,miss} &> & 80 \,\text{GeV} , M_{T,2}  >   120 \,\text{GeV}.
    \label{cuts3}
\end{eqnarray}  

\begin{figure}[h!]
\centerline{
\includegraphics[width=8.2cm]{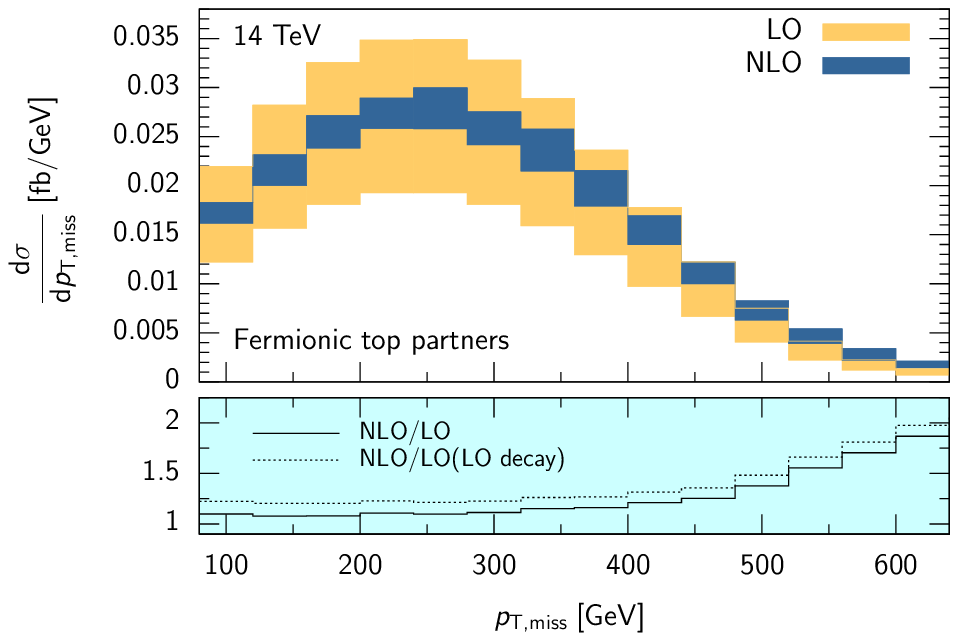}
\includegraphics[width=8.2cm]{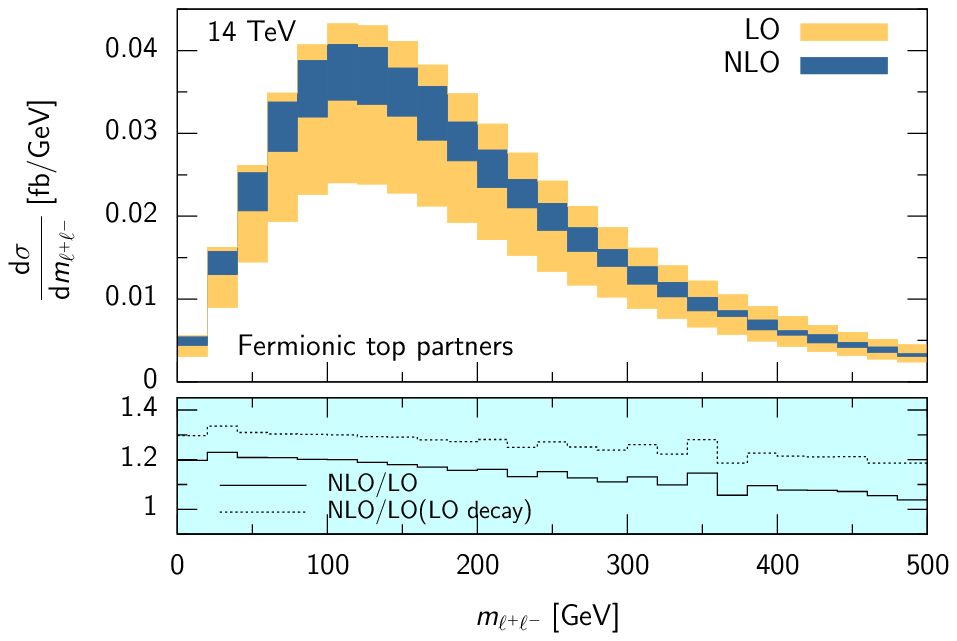}}
\caption{Distributions for fermionic top partners with $(m_T,m_{A_0}) = (600\text{GeV},50\text{GeV})$ at a 14 TeV LHC.  The distributions shown are: the missing transverse momentum and the dileptonic invariant mass.}
\label{fig:tprimes.LHC14.600-50.semi1}
\end{figure}

\begin{figure}[h!]
\centerline{
\includegraphics[width=8.2cm]{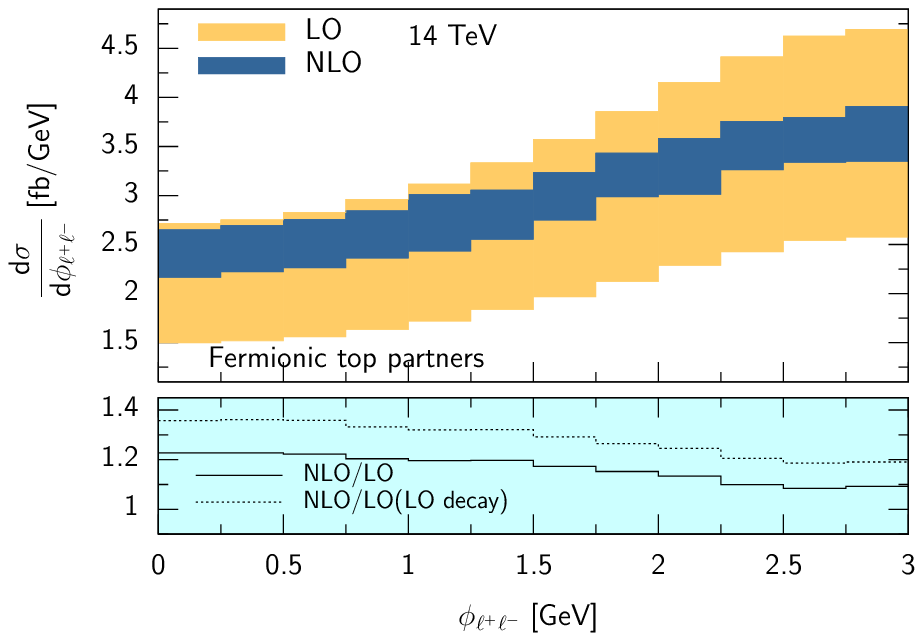}
\includegraphics[width=8.2cm]{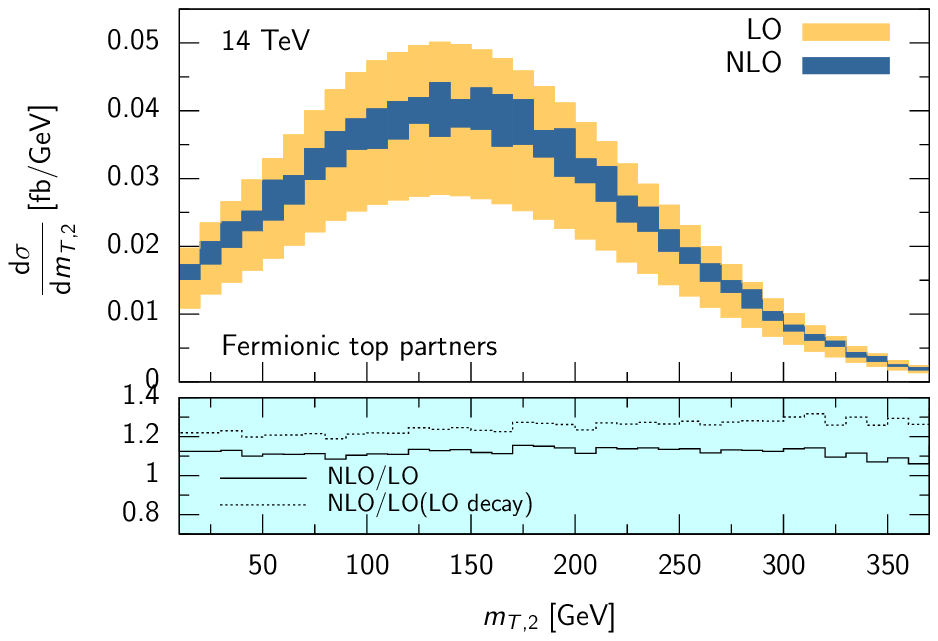}
}
\caption{Distributions for fermionic top partners with $(m_T,m_{A_0}) = (600\text{GeV},50\text{GeV})$ at a 14 TeV LHC.  The distributions shown are the transverse-plane opening angle between the two leptons and $m_{T,2}$. }
\label{fig:tprimes.LHC14.600-50.semi2}
\end{figure}

We also show the result of imposing an additional cut $m_{T,2} > 120$ GeV, which helps reduce the dileptonic $t\bar{t}$ background.  
In both cases the acceptance changes by roughly 10\% in going from LO to NLO.  This behavior is different than that of the semi-leptonic final state studied previously.  QCD corrections do not significantly change the kinematics of the dileptonic mode, and the incorporation of higher-order QCD into the experimental analysis by an overall rescaling works reasonably well. 
Finally, we show in Figs.~\ref{fig:tprimes.LHC14.600-50.semi1} and~\ref{fig:tprimes.LHC14.600-50.semi2} the $p_{T,miss}$, $m_{l^+l^-}$, $\phi_{l^+l^-}$, and $m_{T,2}$ distributions for a fermionic top partner.  The pattern of corrections is similar to that for the scalar partner at 14 TeV.  The $K$-factor for the $p_{T,miss}$ distribution increases as a function of $p_{T,miss}$.  The $m_{l^+l^-}$ and $\phi_{l^+l^-}$ distributions are slightly shifted by QCD corrections, while the $m_{T,2}$ distribution receives a flat correction.  The QCD corrections are in general smaller than for the scalar partner.  Although neglecting the QCD corrections in the decay again induces roughly 10\% errors in the theoretical predictions, this introduces a larger relative error due to the smaller overall $K$-factor.  The size of the $K$-factor may be mis-estimated by up to a factor of two if QCD corrections in the decay are not included.  The cross sections before and after cuts, as well as the acceptances, are shown in Table~\ref{table:14TeVtprime}.  As for the scalar top partner in the dileptonic mode, the acceptance is stable when going from LO to NLO.  The scale dependence reduces from approximately $\pm 30\%$ at LO to $\pm 8\%$ at NLO, independent of whether cuts are imposed.


\begin{table}[htbp]
\begin{center}
\begin{tabular}{| c || c | c | c || c | c |}
\hline
 & $\sigma_{total}$ (fb)& $\sigma_{cut}$ (fb) & acceptance & $\sigma_{cuts+m_{T,2} \, cut}$ (fb) & acceptance \\ \hline \hline
LO & $14.05^{+5.14}_{-3.48}$ & $8.36^{+3.06}_{-2.07}$  & $0.59^{+0}_{-0}$ & $4.90^{+1.79}_{-1.22}$ & $0.35^{+0}_{-0.01}$   \\ \hline
NLO & $16.8^{+1.2}_{-1.8}$  & $9.64^{+0.62}_{-0.97}$  & $0.57^{0.01}_{-0}$  & $5.54^{+0.26}_{-0.53}$ & $0.33^{+0.01}_{-0}$   \\ \hline
\end{tabular}
\end{center}
\caption{Inclusive cross section, the cross section after the experimental cuts shown in Eq.~(\ref{cuts}), and the acceptance for a fermionic top partner, at LO and NLO in QCD, for a 14 TeV LHC. The right part of the table shows the cross section and acceptance after also including a cut on $m_{T,2}$}
\label{table:14TeVtprime}
\end{table}

\subsection{Discriminating scalar and fermionic top partners}

\begin{figure}[h!]
\centerline{
\includegraphics[width=10cm]{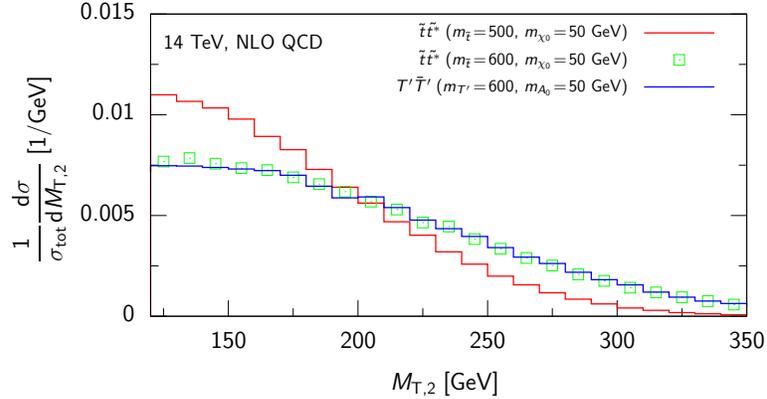}}
\caption{Normalized $m_{T,2}$ distributions for the dileptonic channel at a 14 TeV LHC, for both scalar and 
fermionic top partners, for several mass choices.}
\label{fig:shape-LHC14-dile-mT2}
\end{figure}

\begin{figure}[h!]
\centerline{
\includegraphics[width=10cm]{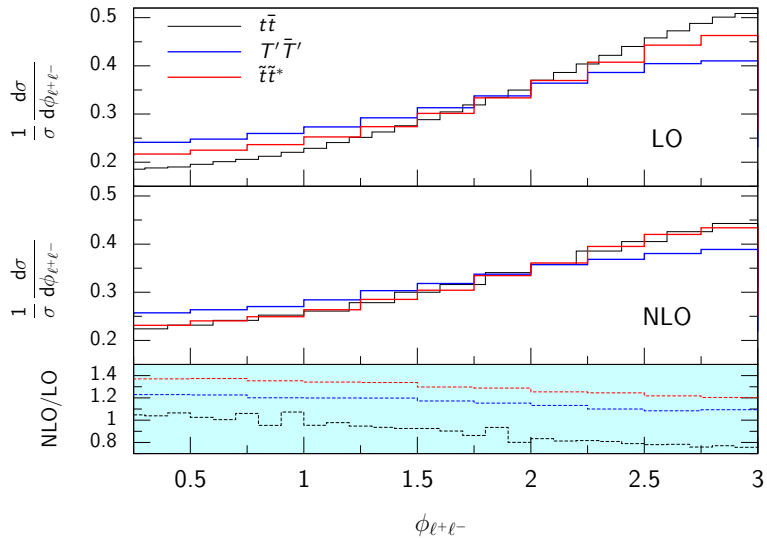}}
\caption{Normalized $\phi_{ll}$ distributions for the dileptonic channel at a 14 TeV LHC, for both scalar and 
fermionic top partners assuming $m_T=600$ GeV and $m_{A_0}=50$ GeV, and for the top-quark background.  Both LO and NLO distributions are shown.}
\label{fig:shape-LHC14-dile-dPhi}
\end{figure}

In this section we study the discriminating power of different variables to the spin of the top partner.  We consider the dileptonic channel at a 14 TeV LHC, and focus on the use of normalized distributions to probe the spin, since the inclusive cross section depends on the unknown overall coupling strength.  We begin by considering the $m_{T,2}$ distribution in Fig.~\ref{fig:shape-LHC14-dile-mT2}.  Two different mass choices for the scalar partner are shown, $m_T=500$ GeV and $m_T=600$ GeV.  For the fermionic partner, we set $m_T=600$ GeV.  Comparison of the two scalar mass points shows that the $m_{T,2}$ distribution has significant sensitivity to the mass parameters.  However, it does not have sensitivity to the partner spin; the scalar and fermionic partner curves for $m_T=600$ GeV lie almost on top of each other.  We next consider $\phi_{l^+l^-}$, this time setting $m_T=600$ GeV for both cases.  These distributions, together with that for the top-quark background, are shown in Fig.~\ref{fig:shape-LHC14-dile-dPhi}.  An interesting difference exists between the distributions at LO and NLO in QCD.  At LO, all three distributions (scalar, fermion, and background) show distinct shape differences.  However, the NLO corrections shift the top-quark background to coincide with the scalar-partner distribution.  Since the $t\bar{t}$ background is large, this shape change must be accounted for in analyses.  The fermionic partner distribution is flatter than the scalar distribution, and discrimination between the two spins should be possible using this distribution.

\section{Conclusions}
\label{sec:conc}

In this paper we have presented a detailed study of the $t\bar{t} + E_{T,miss}$ signature arising from the pair production of top-quark partners at the LHC.  We have considered scalar and fermionic top partner production through next-to-leading order in perturbative QCD.  Higher-order corrections have been included consistently throughout the entire decay chain in the narrow-width approximation.  We presented numerical results for both the semi-leptonic and fully-leptonic decays of the $t\bar{t}$ pair, and have considered a host of kinematic variables that either distinguish signal from background, or assist in discriminating between possibilities for the top-partner spin.

There are several conclusions that can be drawn from our study.  The first is that in general, leading-order plus parton-shower simulations do not provide a good framework for modeling new physics signals.  In our case study in the semil-leptonic channel at an 8 TeV LHC, the acceptance obtained using the default Pythia settings included with Madgraph differed from the actual NLO value by nearly a factor of two.  However, we found that a leading-order merged sample containing an additional hard jet reproduced the NLO prediction for the acceptance.  Both the merged sample and the NLO prediction seem to suitably model the signal shape, although only the NLO result correctly obtains the normalization and reduces the scale uncertainty.  We note that the tuned Herwig simulation used in the current ATLAS searches for stop-quark pair production correctly produces the next-to-leading order distribution shapes, indicating that the acceptances used in this experimental analysis are correct.

Another conclusion we draw from our results is that the scale variation of the inclusive cross section does not accurately reflect the uncertainty in the theoretical cross section after cuts are imposed.  In our study this was most striking when we considered a compressed spectrum at the 8 TeV LHC; the scale uncertainty after cuts was almost a factor of two larger than the variation of the inclusive cross section.  The scale variation of the cross section after cuts should be used as the theoretical systematic error when setting exclusion limits, and we encourage the experimental collaborations to quantify the effect by revising their error estimate.

Finally, we compared several distributions that could potentially provide a handle on determining the spin of the top-quark partner in the dileptonic mode, including $m_{T,2}$ and $\phi_{l^+l^-}$.  We found that $m_{T,2}$ does not distinguish between scalar and fermionic top partners if they had equal masses.  We found that $\phi_{l^+l^-}$ did discriminate between the two possibilities, but that it is important to use NLO QCD predictions in this analysis.  The NLO corrections shift the top-quark background to coincide with the scalar-partner distribution, and the determination of signal over background in this analysis requires an NLO QCD prediction.

%
%
%
%
%

\vspace{.5cm}
\medskip
\noindent
{\bf Acknowledgments:} 
\vspace{0.5cm}

We thank Till Eifert for useful communications and input. We thank the Erwin Schr\"odinger International Institute for Mathematical Physics (ESI) and the Aspen Center for Physics for kind hospitality while this work was being completed. 
This research is supported by the US DOE under contract DE-AC02-06CH11357.
The submitted manuscript has been created by UChicago Argonne, LLC, Operator of Argonne National Laboratory (Argonne). Argonne, a U.S. 
Department of Energy Office of Science laboratory, is operated under Contract No. DE-AC02-06CH11357. The U.S. Government retains for 
itself, and others acting on its behalf, a paid-up nonexclusive, irrevocable worldwide license in said article to reproduce, prepare 
derivative works, distribute copies to the public, and perform publicly and display publicly, by or on behalf of the Government.

\end{document}